\def\diff{\mathrm{d}}

\def\n{\mathrm{n}}
\def\p{\mathrm{p}}
\def\q{\mathrm{q}}
\def\s{\mathrm{s}}
\def\v{\mathrm{v}}

\def\A{\mathrm{A}}
\def\B{\mathrm{B}}
\def\C{\mathrm{C}}
\def\D{\mathrm{D}}
\def\Fe{\mathrm{F}}	
\def\UU{\mathrm{UU}}
\def\NN{\mathrm{NN}}

\def\vecr{\mathbf{r}}
\def\vecp{\mathbf{p}}
\def\veck{\mathbf{k}}

\def\Ntest{N_{\textrm{test}}}

\def\eF{\epsilon_{\textrm{F}}}
\def\vF{v_{\textrm{F}}}

\def\bbsty#1#2#3#4{#4 #1 {\bf #2} #3}	

\documentclass[a4paper]{jpconf}
\usepackage{graphicx}\usepackage{epsfig}
\usepackage{amsmath}\usepackage{amssymb}\usepackage{amsfonts}
\usepackage{bm}
\usepackage{exscale}
\usepackage{dcolumn}    
\usepackage{color}
\usepackage{soul}
\usepackage{ulem}

\begin{document}
\title{Cluster formation in nuclear reactions from mean-field inhomogeneities}

\author{Paolo Napolitani$^1$, Maria Colonna$^2$ and Carlo Mancini-Terracciano$^3$}

\address{$^1$ IPN, CNRS/IN2P3, Universit\'e Paris-Sud, Universit\'e Paris-Saclay, 91406 Orsay, France}
\address{$^2$ INFN-LNS, Laboratori Nazionali del Sud, 95123 Catania, Italy}
\address{$^3$ INFN, Sezione di Roma, Rome, Italy}
%
%

\begin{abstract}
Perturbing fluids of neutrons and protons (nuclear matter) may lead, as the most catastrophic effect, to the rearrangement of the fluid into clusters of nucleons.
A similar process may occur in a single atomic nucleus undergoing a violent perturbation, like in heavy-ion collisions tracked in particle accelerators at around 30 to 50 MeV per nucleon: in this conditions, after the initial collision shock, the nucleus expands and then clusterises into several smaller nuclear fragments.

Microscopically, when violent perturbation are applied to nuclear matter, a process of clusterisation arises from the combination of several fluctuation modes of large-amplitude where neutrons and protons may oscillate in phase or out of phase.
The imposed perturbation leads to conditions of instability, the wavelengths which are the most amplified have sizes comparable to small atomic nuclei.
We found that these conditions, explored in heavy-ion collisions, correspond to the splitting of a nucleus into fragments ranging from Oxygen to Neon in a time interval shorter than one zeptosecond (10$^{-21}$s).
From  the out-of-phase oscillations of neutrons and protons another property arises, the smaller fragments belonging to a more volatile phase get more neutron enriched: in the heavy-ion collision case this process, called distillation, reflects in the isotopic distributions of the fragments.

The resulting dynamical description of heavy-ion collisions is an improvement with respect to more usual statistical approaches, based on the equilibrium assumption. It allows in fact to characterise also the very fast early stages of the collision process which are out of equilibrium.
Such dynamical description is the core of the Boltzmann-Langevin One Body (BLOB) model, which in its latest development unifies in a common approach the description of fluctuations in nuclear matter, and a predictive description of the disintegration of nuclei into nuclear fragments.
After a theoretical introduction, a few practical examples will be illustrated.

	This paper resumes the extended analysis of fluctuations in nuclear matter of ref.~\cite{Napolitani2017} and briefly reviews applications to heavy-ion collisions.

\end{abstract}

\section{Introduction}\vspace{.75ex}

The most catastrophic process which can occur in a nuclear complex is its splitting into clusters and fragments when undergoing a violent external action.
	We want to address this process, which can be probed in a dissipative heavy-ion collision, from the point of view of dynamics, moving from nuclear matter to nuclei, which are finite open self-bound systems.
	Nuclear clusterisation may appear in various forms.
	Exotic topologies in nuclear astrophysics (stellar matter), nuclear states explored in low-energy reactions from the recombination and vibrations of existing cluster structures, low-density regions of the equation of state landscape, where clusterisation in intermediate-mass fragments and clusters may arise and can be explained in violent nuclear reactions.
	Finally, clusterisation characterises in general Fermi liquids as arising from ripples produced by phase-space fluctuations~\cite{Pines1966}.

	We constructed a microscopic dynamical framework from applying the theory of Fermi liquids to clusterisation in nuclei (see ref.~\cite{Napolitani2017} for a more extended and detailed discussion); within this framework, we explore how the clusterisation progresses from zero-sound propagation.
	Two avenues are mostly followed to describe nuclear processes, molecular dynamics and mean-field approaches~\cite{Ono2006_WCI}.
In mean-field approaches particles evolve independently in their own self-consistent mean-field.
	Such treatment, while well suited to describe the collective behaviour, can not handle clusterisation in its pure mean-field form.
	On the other hand, molecular-dynamics approaches rely on a description in terms of product wave functions.
	The independent-particle scheme (i.e. mean field) is applied to nucleonic single-particle wavefunctions $\psi_i$, so that many-body correlations in both mean field and scattering can be achieved from the localisation of $\psi_i$ (a coherent-state subspace is used in FMD~\cite{Feldmeier1990_1995} and an even stronger localisation is imposed in AMD~\cite{Ono1992,Ono2016} by also fixing the widths of $\psi_i$).
	Such treatment is successful for final-state correlations, but it may approximate collective behaviour and 0-sound propagation.
	
	To profit from a well suited description of 0-sound propagation, we follow thereafter a mean-field approach where, in order to describe clusterisation, extensions to handle large-amplitude dynamics should be explicitly introduced.
	In particular, correlations beyond the level of kinetic equations are needed or, in terms of BBGKY hierarchy~\cite{Balescu1976}, upper orders beyond two-body correlations should be recovered in order to access highly non-linear regimes.
	A technique to obtain such result in an approximated form consists in applying two treatments in parallel: first, introducing nucleon-nucleon correlations continuously, second, handling a stochastic ensemble of several mean-field trajectories to exploit the introduced correlations.
	In this case, a mean-field trajectory $\rho_1$ taken at a given time, would split at a successive time into subensembles $\rho_1^{(n)}$:
\begin{equation}
	\rho_1 \longrightarrow \{ \rho_1^{(n)}; n=1, \dots, \textrm{subens.} \}   \;.
\end{equation}

	In practice, at the level of 
kinetic equations,
not only collisional correlations should be introduced, but also a term of vanishing mean which injects fluctuations around the collision integral intermittently in time.
	Such term can be exploited as a stochastic source to revive fluctuations all along a dissipative process.
	Adding the collisional correlations introduced above and fluctuations to the mean-field produces a scheme which resembles stochastic TDHF~\cite{Ayik1988,Lacombe2016}:
\begin{equation}
	i\hbar\frac{\partial\rho_1^{(n)}}{\partial t} \approx [k_1^{(n)}+V_1^{(n)} , \rho_1^{(n)}] 
		+ \bar{I}_{\textrm{coll}}^{(n)} + \delta I_{\textrm{coll}}^{(n)}   \;,
\label{eq:STDHF}
\end{equation}
	where  $\bar{I}_{\textrm{coll}}^{(n)}$ and  $\delta I_{\textrm{coll}}^{(n)}$
are the average collision contribution and a continuous source of fluctuation seeds, respectively.
	The corresponding Wigner transform yields the Boltzmann-Langevin (BL) equation~\cite{Reinhard1992}, in terms of an ensemble of distribution function $f^{n}$,
\begin{equation}
	\frac{\partial f^{(n)}}{\partial t} = \{h^{(n)} , f^{(n)}\} 
		+ I_{\textrm{UU}}^{(n)} + \delta I_{\textrm{UU}}^{(n)}   \;,
\label{eq:SMF}
\end{equation}
where the ensemble $f^{(n)}$ replaces the ensemble of Slater determinants in Eq.~(\ref{eq:STDHF}) and corresponds to a Fermi statistics at equilibrium, $h^{(n)}$ is the effective Hamiltonian acting on $f^{(n)}$, and the residual contributions of Eq.~(\ref{eq:STDHF}) are replaced by corresponding Uehling-Uhlenbeck (UU) terms.

	We focus on the BL equation and we solve it in full phase space through the BLOB approach~\cite{Napolitani2013,Napolitani2012,Napolitani2017}, where nucleon-nucleon correlations are introduced by constructing a fluctuating collision term which acts on extended equal-isospin phase-space portions for each single in-medium collision:
\begin{equation}
	\frac{\partial f^{(n)}}{\partial t} - \{h^{(n)} , f^{(n)}\} 
		= I_{\textrm{UU}}^{(n)} + \delta I_{\textrm{UU}}^{(n)}
	= g\int\frac{\diff\vecp_b}{h^3}\,
	\int
	W({\scriptstyle\A\B\leftrightarrow\C\D})\;
	F({\scriptstyle\A\B\rightarrow\C\D})\;
	\diff\Omega\;,
\label{eq:BLOB}
\end{equation}
where $g$ is the degeneracy factor, $W$ is the transition rate in terms of relative velocity between the two colliding phase-space portions, and $F$ handles the Pauli blocking of initial and final states over their full phase-space extensions.
	The extended size of phase-space portions involved in the scattering in Eq.~(\ref{eq:BLOB}) should be large enough so that the occupancy variance in $h^3$ cells is equal to $f(1-f)$, which corresponds to the scattering of two nucleons~\cite{Rizzo2008}.
	For comparison, we consider a second simplified approach to solve the BL equation, Eq.~(\ref{eq:SMF}), based on the SMF~\cite{Colonna1998} treatment, where fluctuations are injected from an external stochastic contribution $U_{\textrm{ext}}$ and projected on spacial density.

	In both cases, in all following calculations, a simplified SKM* effective interaction, with momentum dependence omitted, is used~\cite{Guarnera1996,Baran2005}.
 	A soft isoscalar equation of state with a compressibility modulus $k=200$MeV is used.
	For test, both asy-stiff (linear) and asy-soft (quadratic) parametrisations are used for the symmetry energy.
	All nuclear-matter calculations employ a periodic box of edge $L=39$fm, and the system is initialised with a Fermi-Dirac distribution at a temperature $T=3$MeV.
	In the collision term, either a free or a constant nucleon-nucleon cross section $\sigma_\NN$ is used.

	Before carrying on a study on heavy-ion collisions, we require that fluctuation amplitudes are consistent with analytic expectations from Fermi liquids: for this purpose, we study nuclear matter in initially homogeneous conditions.

\section{Fluctuations in two-component nuclear matter  \label{sec_NM}}\vspace{.75ex}

	While unperturbed, the dynamics of a periodic portion of uniform nuclear matter at low temperature (so that the collision term $I_\UU$ in the analytic description can be neglected) would evolve along a mean trajectory $f^0$.
    Let us introduce perturbations $\delta f^q \ll f^0$ in two forms: either neutrons and protons move in phase (isoscalar perturbation, indexed with $\q=\s$), so that
$\delta f^\q = \delta f^\s = (f_\n-f_\n^0)+(f_\p-f_\p^0)$,
or out of phase (isovector perturbation, indexed with $\q=\v$), so that
$\delta f^\q = \delta f^\v = (f_\n-f_\n^0)-(f_\p-f_\p^0)$ .
    In a periodic box, this action introduces fluctuations $\rho_k^\q$ associated with plane waves of wave number $\veck$, with a corresponding equilibrium variance $(\sigma_k^\q)^2$ (or intensity of response), and with an equilibrium variance of spacial density correlations $(\sigma_{\rho^\q})^2$
(this latter variance is obtained from the former through an inverse Fourier transform).

    If the Boltzmann-Langevin equation is applied to the disturbance $\delta f^\q$, the following scheme can be established.

\begin{itemize}

\item    First, if fluctuations are of stable nature, the two equilibrium variances $(\sigma_k^\q)^2$ and $(\sigma_{\rho^\q})^2$ can be related to the free-energy density curvature $F^\q(k)$ through the fluctuation-dissipation theorem:
\begin{equation}
	(\sigma_k^\q)^2 = \frac{T}{F^\q(k)}\;; \;\;\; (\sigma_{\rho^\q})^2 = \frac{T}{\Delta V}\Big\langle\frac{1}{F^\q(k)}\Big\rangle_\veck
	\;,
\label{eq:fluctuation_dissipation}
\end{equation}
where $T$ is the temperature and $\Delta V$ a volume cell in configuration space where $(\sigma_{\rho^\q})^2$ is calculated.

An application of such scheme to nuclear matter is well suited to investigate the 
equilibrium variance of spacial isovector density correlations $(\sigma_{\rho^\v})^2$
as a function of the symmetry energy.
From an analogous application to open system, the isotopic distributions of clusters emerging from density ripples can be analysed.

\item    Second, in unstable conditions, both initial fluctuation seeds and intermittent fluctuation seeds injected at later times can yield an exponential growth of the intensity of response $(\sigma_k^\q)^2$ over a growth time $\tau_k$~\cite{Colonna1993,Colonna1994_a}.

An application of such scheme to nuclear matter is well suited to sample zero-sound propagation and investigate the instability growth rates $\Gamma_k=1/\tau_k$.
Analogous conditions in open system are suited to investigate the arising of clusterisation.

\end{itemize}


In the following, these steps are attended to.

\subsection{Relating isovector fluctuations to the symmetry energy  \label{sec_iv}}\vspace{.75ex}

To undertake a simulation on the propagation of isovector fluctuations, it is convenient to isolate them from the overwhelming effect of isoscalar fluctuations; those latter would in fact completely dominate the dynamics with their larger amplitude.
Scalar terms are therefore suppressed in the potential so that stable conditions are imposed at any density $\rho^0$.

In this case, the fluctuation--dissipation relation Eq.~(\ref{eq:fluctuation_dissipation}) takes a form where the isovector-density variance is related to an effective symmetry energy, proportional to the symmetry energy $E_{\textrm{sym}}$ at zero temperature~\cite{Colonna2013}.

\begin{equation}
	(\sigma_{\rho^\v})^2 = \frac{T}{\Delta V}\Big\langle\frac{1}{F^\v(k)}\Big\rangle_\veck\;\longrightarrow \;\;\; 
	F_{\textrm{eff}}^\v = 
	\frac{T}{2\Delta V} \frac{\rho^0}{(\sigma_{\rho^\v})^2} = \frac{T}{2\Delta V} \frac{\rho^0}{\langle [\delta\rho_n(\vecr)-\delta\rho_p(\vecr)]^2 \rangle} 
\;\;\propto\;E_{\textrm{sym}}(\rho^0)
	\;,
\label{eq:fluctuation_dissipation_iv}
\end{equation}

This relation can be solved numerically by calculating the isovector variance 
$(\sigma_{\rho^\v})^2$ in cells of edge size $l$, where $l$ should be chosen large enough to minimise surface effects and better sample the volume symmetry energy (this value is found around $l=2$fm).

%
%
\begin{figure}[h]
\includegraphics[width=.550\textwidth]{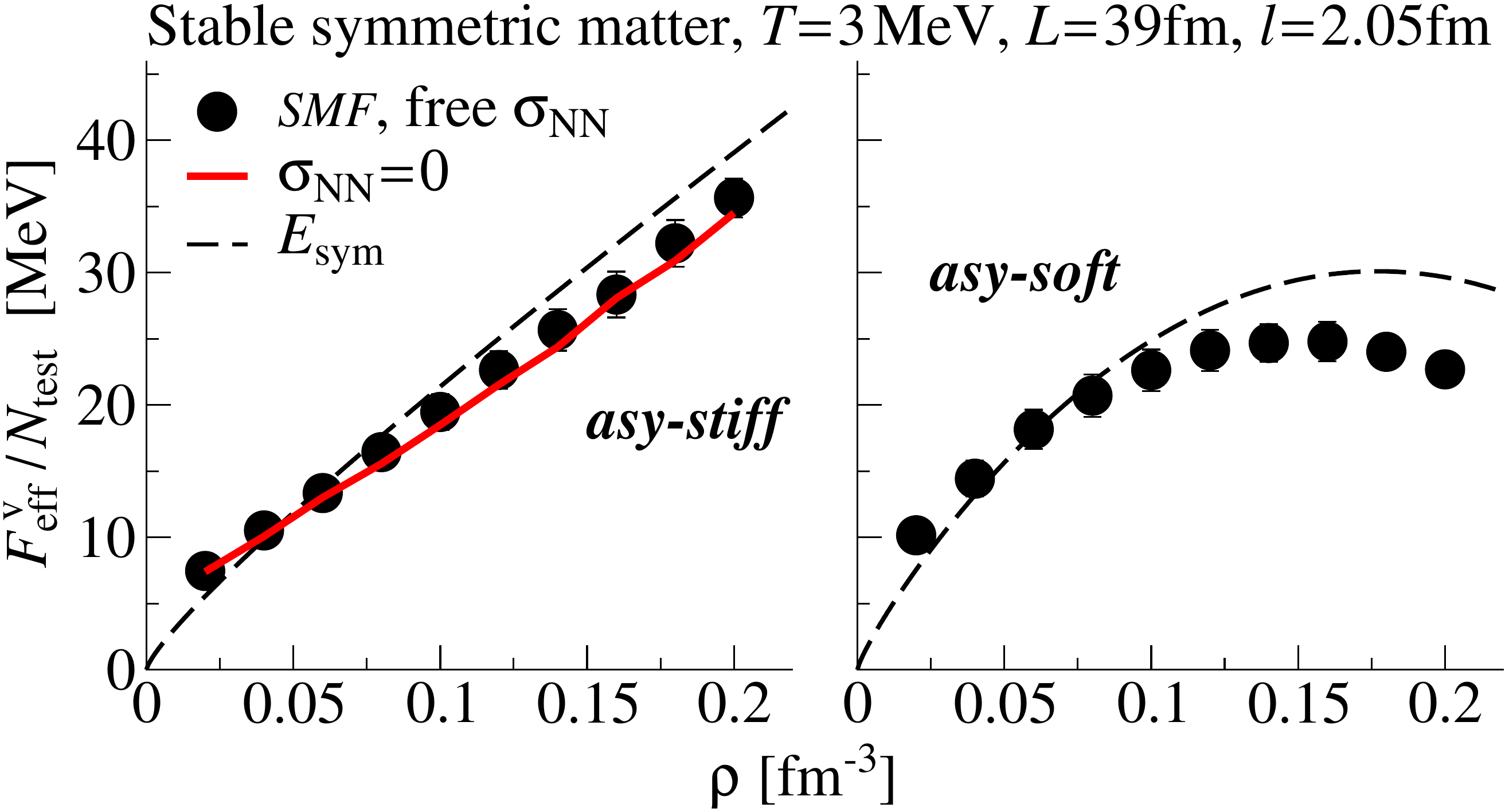}\hspace{2pc}%
\begin{minipage}[b]{.395\textwidth}\caption{\label{fig_Fivrho_Tcorr_SMF}
SMF calculation of the effective symmetry energy, scaled by $\Ntest$, compared to the analytic expression of the symmetry energy for different system densities.
The collision term is either activated (and a free $\sigma_\NN$ is used) or suppressed
($\sigma_\NN=0$).
Two parametrisations of the symmetry energy, either asy-stiff (left panel), or asy-soft (right panel) are used.
}
\end{minipage}
\end{figure}

	In a first calculation, shown in Fig.~\ref{fig_Fivrho_Tcorr_SMF}, where the simplified SMF approach to solve the BL equation is used, even though the evolution of the effective symmetry energy $F_{\textrm{eff}}^\v$ as a function of density reproduces correctly the shape of the analytic expression for the symmetry energy $E_{\textrm{sym}}$, the former and the latter differ of a factor equal to the number of test particles per nucleon $\Ntest$ employed in the numerical sampling of the mean field:
\begin{equation}
	F_{\textrm{eff}}^\v \approx \Ntest E_{\textrm{sym}}
	\;.
\label{eq:Ntest_scaling}
\end{equation}
	The same result is obtained with or without the contribution of the residual term $I_\UU$, indicating that explicit isovector terms are missing in the isovector channel in building fluctuations.
	This is not surprising because the residual term is not defined to build nucleon-nucleon correlation and the corresponding fluctuations.
	Those latter, even though reflecting the potential employed in the calculation, develop from initial fluctuation seeds which are related to the numerical noise in the sampling of the mean-field, which, on its turn, is related to $\Ntest$.

%
%
\begin{figure}[h]
\includegraphics[width=.685\textwidth]{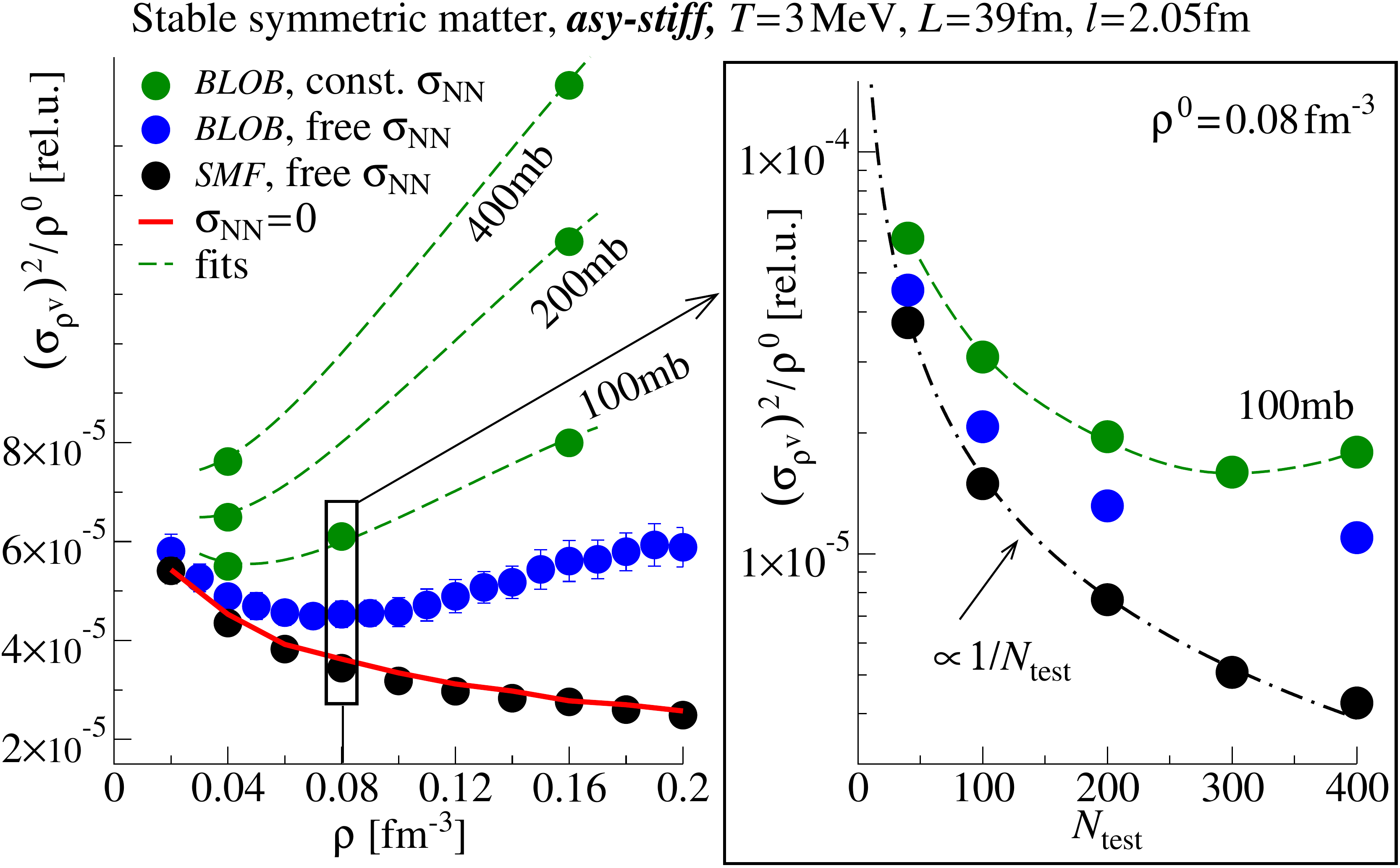}\hspace{2pc}%
\begin{minipage}[b]{.26\textwidth}\caption{\label{fig_Ntest_scaling}
Isovector variance at equilibrium as a function of the system density for an asy-stiff parametrisation of $E_{\textrm{sym}}$.
Calculated with BLOB, with different values of $\sigma_\NN$, and compared with SMF, with the collision term either activated (with a free $\sigma_\NN$) or suppressed.
For the system density 0.08fm$^{-3}$, the inset shows the effect of varying $\Ntest$ (see text).
}
\end{minipage}
\end{figure}
	A second calculation, shown in Fig.~\ref{fig_Ntest_scaling}, employs the BLOB approach, where the residual term is designed to explicitly introduce phase-space fluctuations from nucleon-nucleon correlations.
	In this case, the isovector variance is enhanced, but equilibrated nuclear matter is not a favourable condition to completely recover the large isovector variance of the analytic expectation.
	In particular, at small density $\rho^0$ collisional correlations are ineffective, at larger density collisions become rare and can hardly revive fluctuations and, in addition, the same noise issue already observed in the SMF calculation is acting as a smearing contribution~\cite{Colonna1993} with a dependence on $\Ntest$.

	For test purposes, it is instructive to appreciate the effect of increasing the nucleon-nucleon cross section and, as a consequence, intensifying the nucleon-nucleon collision rate.
	This action enhances the effect of the BLOB fluctuating term in reviving nucleon-nucleon collisions, so as to prevail over the smearing effect of the mean-field noise.

	A more consistent (but numerically costly) approach is reducing the numerical noise by refining the mean-field paving: this action, consisting in simply increasing $\Ntest$,  also improves the results with respect to the SMF approach.

	From these results and trends, we can expect that the limits imposed by the conditions of equilibrated nuclear matter could be actually overcame in out-of-equilibrium conditions, where nucleon-nucleon collision rates are definitely higher: this situation corresponds to the early stages of heavy-ion collisions, which are investigated in Sec.~\ref{sec_densityripples}.

\subsection{Relating zero-sound propagation to instability growth}\vspace{.75ex}

	To undertake a simulation of zero-sound propagation of collective modes in nuclear matter, the system should be prepared as initially homogeneous and at low temperature.
	If the temperature increases, a transition from zero to first sound may in fact occur~\cite{Larionov2000,Kolomietz1996}.
	If unstable conditions are investigated, an amplification of the fluctuation amplitude is expected to trigger a catastrophic process.
	Early times are therefore better suited to compare to analytic expectations.
 	If the zero-sound propagation is correctly sampled, isoscalar fluctuations should develop spontaneously with the correct amplitude, inducing the arising of a mottling pattern.

	The analytic expectation can be defined in a linear-response approximation, assuming small deviations from a mean trajectory $f^0$; in this case, from the linearised Vlasov equation (where residual terms are suppressed) expressed as a function of the disturbance wave number $k$ and frequencies $\omega_k$, the corresponding dispersion relation~\cite{Landau1957,Kalatnikov1958} can be extracted by applying self-consistency, as follows:
\begin{equation}
	\omega_k f_k + \veck\cdot\frac{\vecp}{m}f_k - \frac{\partial f^0}{\partial \epsilon}\frac{\partial U_k}{\partial \rho} \veck\cdot\frac{\vecp}{m} \rho_k = 0
	\;\;\longrightarrow\;\;
	1 = \frac{g}{h^3}  \frac{\partial U_k}{\partial\rho} \int \frac{\partial f^0}{\partial\epsilon}  \frac{\veck\cdot\vecp /m}{\omega_k+\veck\cdot\vecp /m} \diff\vecp
	\;.
\label{eq:linearised_Vlasov}
\end{equation}
	At zero temperature, the eigenmodes $f_k$ in the linearised Vlasov equation (left side of the arrow) depend on states near the Fermi level $\epsilon_\Fe$ and the integral in the dispersion relation (right side of the arrow) is restricted to the Fermi surface so that, introducing the Landau parameter 
$F_0(\veck) = (3/2)(\rho^0/\eF)\partial_\rho U_k$, the dispersion relation can be written in terms of sound velocity $s=\omega_k / (k v_\Fe)$~\cite{Colonna1994_a,Chomaz2004}
\begin{equation}
	1 + \frac{1}{F_0} = \frac{s}{2} \textrm{ln} \left(\frac{s+1}{s-1} \right)
	\;,
\label{eq:disprel_T0}
\end{equation}

%
%
\begin{figure}[t!]
\includegraphics[width=1\textwidth]{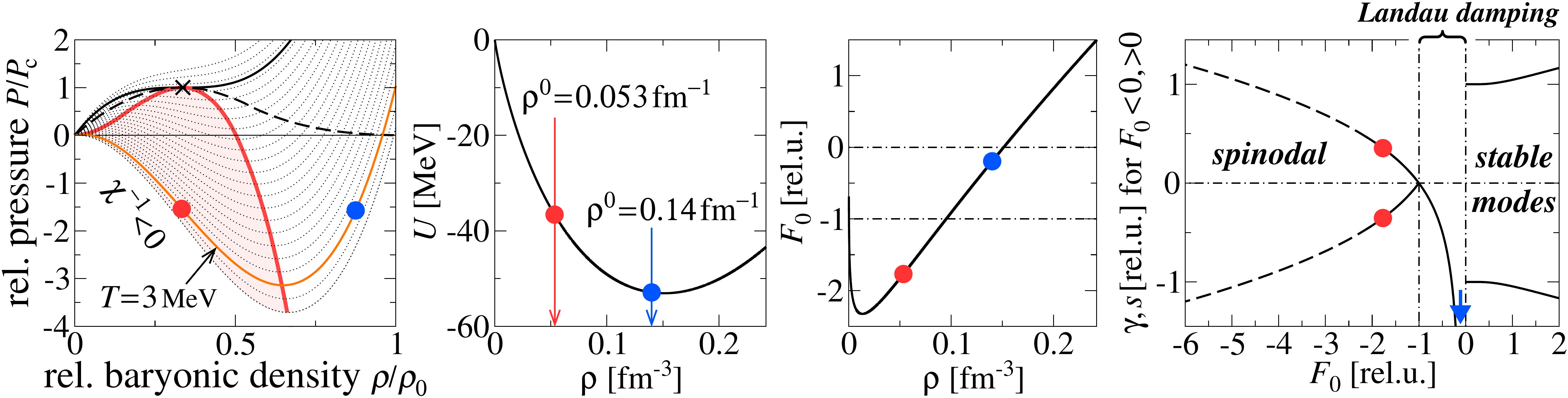}\hspace{2pc}%
\begin{minipage}[b]{1\textwidth}\caption{\label{fig_EOS}
Two conditions, corresponding to the spinodal instability ($\rho^0=0.053$fm$^{-3}$) and Landau damping ($\rho^0=0.14$fm$^{-3}$) are indicated with dots in a sequence of four plots. 
From left to right: 
the equation-of-state landscape;
the isoscalar mean-field potential as a function of the density;
the Landau parameter $F_0$ as a function of the density;
real (for $F_0>0$) and imaginary (for $F_0<0$)  roots of the dispersion relation as a function of the Landau parameter $F_0$.
}
\end{minipage}
\end{figure}
%
%
%
\begin{figure}[b!]
\includegraphics[width=1\textwidth]{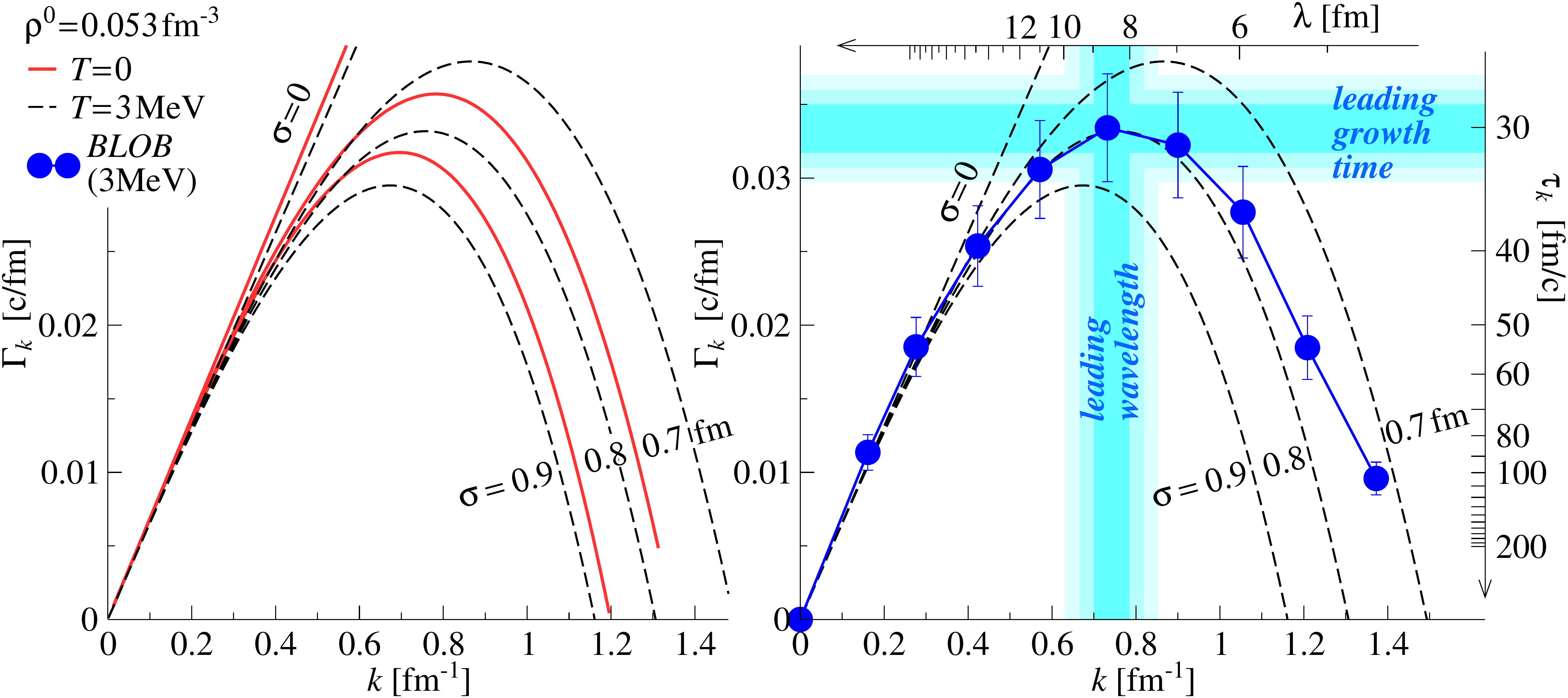}\hspace{2pc}%
\begin{minipage}[b]{1\textwidth}\caption{\label{fig_disprel_BLOB}
growth rate $\Gamma_k$ (or growth time $\tau_k$) as a function of wave number $k$ (or wavelength $\lambda$) in nuclear matter at $\rho^0=0.053$fm$^{-3}$.
(Left) Analytic predictions from Eq.~(\ref{eq:disprel_gamma}) for $T=0$ and $T=3$MeV and for different values of the Gaussian smearing factor $\sigma$.
(Right) BLOB calculation from Eq.~(\ref{eq:tau_numerical}) for $T=3$MeV compared to the analytic predictions (the interaction, considering surface contributions, uses a $\sigma$ in the region of 8 to 9fm): the resulting leading wavelength and growth time are indicated.
Uncertainties are evaluated from the linear-response fit. 
}
\end{minipage}
\end{figure}
	As shown in Fig.~\ref{fig_EOS}, instabilities in zero-sound conditions can be tracked by placing the system in a location of the equation-of-state landscape where the incompressibility
$\chi^{-1}\equiv \rho\frac{\partial P}{\partial\rho}$ is negative.
	In these conditions, corresponding to the spinodal instability and to $F_0(k\!=\!0)<-1$, the dispersion relation yields imaginary roots~\cite{Pomeranchuk1959} which, by replacing $s\rightarrow i\gamma$, can be put in the form
\begin{equation}
	1 + \frac{1}{F_0(k)}
	= \gamma\,\textrm{arctan}\frac{1}{\gamma}
	\;\;\longrightarrow\;\;
	|\gamma| = \frac{|\omega_k|}{k\vF} = \frac{1}{\tau_k k \vF}
	\;.
\label{eq:disprel_gamma}
\end{equation}
	Eq.~(\ref{eq:disprel_gamma}) shows that disturbances of wave number $k$ get amplified
with a growth time $\tau_k$ and a corresponding growth rate $\Gamma_k=1/\tau_k$.

	As shown in the analytic calculations of Fig.~\ref{fig_disprel_BLOB} (left panel) for zero temperature, the response intensity at zero sound should present the following evolution of the growth rate $\Gamma_k$ as a function of the $k$ number, or with the corresponding wavelength $\lambda$.
\begin{itemize}
\item For small $k$: $\Gamma_k$ tends to increase (decrease) linearly with $k$ ($\lambda$) because the more matter has to be relocated, the longer time it takes.
\item For large $k$: small wavelengths $\lambda$ are excluded as a function of the interaction range. This ultraviolet cutoff can be described as a Gaussian smearing~\cite{Colonna1994,Kolomirtz1999} $\sigma$ of the mean-field potential in configuration space $U\otimes g(k)$ with $g(k)=\textrm{e}^{-\frac{1}{2} (k\sigma)^2}$ which reduces $\Gamma_k$ with $k$.
\item The combination of these opposite behaviours produces a maximum which corresponds to the fastest growing disturbance, i.e. the leading $k$ mode and leading wavelength. 
\end{itemize}

	To explore the equation of state (especially in applications to heavy-ion collisions), it is convenient to have a finite value of $T$.
	Still, as mentioned, $T$ should be low in order to avoid transitions to first-sound.
	It is possible to introduce a finite temperature $T$ rather than working at $T=0$ through a low-temperature expansion of the chemical potential $\nu$.
	A reduction of $\Gamma_k$ with $T$ is the result of this action.

	These modifications to take into account the ultraviolet cutoff and to introduce a finite temperature impose to replace $F_0(k)$ by an effective Landau parameter~\cite{Colonna1994} $\widetilde{F}_0(k,T) = (\mu(T)/\eF) F_0 g(k)$ in eq.~(\ref{eq:disprel_gamma}) in order to obtain the analytic response in Fig.~\ref{fig_disprel_BLOB} for a finite temperature.

%
%
\begin{figure}[b]
\includegraphics[width=.645\textwidth]{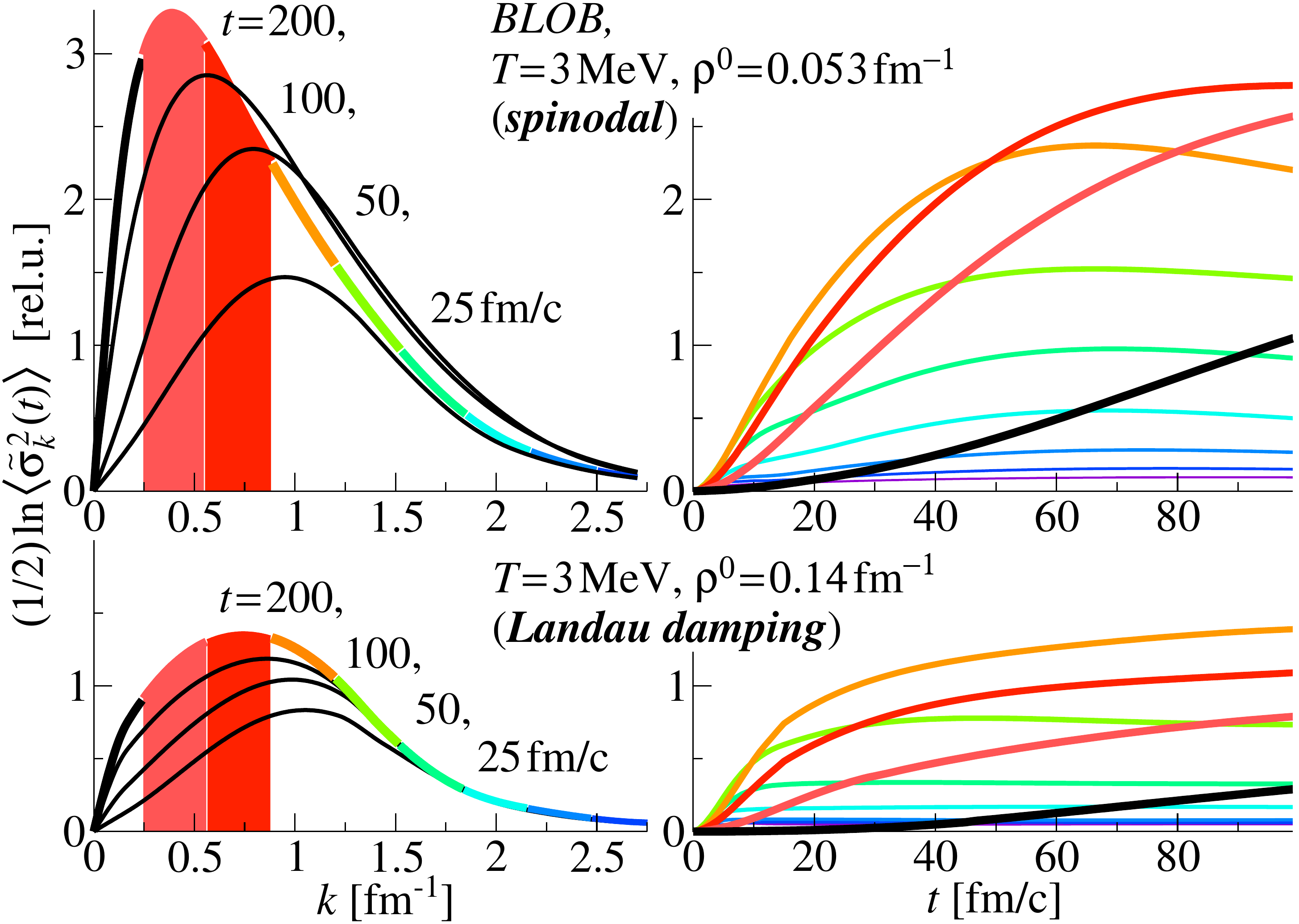}\hspace{2pc}%
\begin{minipage}[b]{.3\textwidth}\caption{\label{fig_growth_modes}
Average response intensity 
as a function of $k$ (left column) and time $t$ (right column), calculated with BLOB in conditions of spinodal-instability (top row) and Landau damping (bottom row).
The different curves in the left column correspond to different time instants, while the different curves in the right column correspond to different $k$ intervals (their colours corresponds to the coloured segments in the left panel).}
\end{minipage}
\end{figure}
	The corresponding numerical calculations 
employ the BLOB approach to extract the response intensity $\sigma_k^2(t)$, i.e. the amplitude of the isoscalar fluctuation of a mode $k$ from the Fourier transform of the space density (see ref.~\cite{Napolitani2017} for details).
	Fig.~\ref{fig_growth_modes} studies the evolution of the ratio $\tilde{\sigma}^2_k(t) = \sigma^2_k(t)/ \sigma^2_k(t=0)$.

	From the analysis of the response intensity we infer some general results which can be transposed from nuclear matter to heavy-ion collisions.
\begin{itemize}
\item in unstable (spinodal) conditions, calculated at $\rho^0=0.053$~fm$^-3$ fluctuations get rapidly amplified as expected.
\item Even outside of the spinodal instability, in calculations at $\rho^0=0.14$~fm$^-3$, which correspond to Landau-damping conditions, the response intensity reaches significant amplitudes.
\item The response intensity saturates rather early due to the combination of large $k$ into small $k$. In heavy-ion collisions, a similar effect manifests in the recombination of small emerging clusters into larger fragments and in an overall mean-field resilience effect (investigated in detail in ref.~\cite{Napolitani2015}).
\end{itemize}

	To the analytic expectation for the growth-rate as a function of $k$, Fig.~\ref{fig_disprel_BLOB} (right panel) overlaps the corresponding numerical calculation obtained with BLOB for the same conditions and interaction properties, when taking the average over several stochastic dynamical paths,
\begin{equation}
\Gamma_k = \frac{1}{2} \frac{\partial}{\partial t} {\textrm ln}\prec \tilde{\sigma}^2_k(t) \succ 
	\;.
\label{eq:tau_numerical}
\end{equation}
	The achieving of a close correspondence between the BLOB calculation and the analytic expectation, as shown in Fig.~\ref{fig_disprel_BLOB} ensures that the model succeeds in describing the fluctuation phenomenology consistently and, in particular, fluctuations develop with the correct amplitude.
	After showing that this correspondence is achieved, some quantitative results can be extracted from the calculation.
	The leading wavelength ranges between 8 to 9~fm. 
	By analogy to nuclear matter, we can expect for heavy-ion collisions that fragments and clusters should arise in the region of Neon. 
	The corresponding separation time in heavy-ion collisions should not only take into account the growth time of the leading $k$ modes, calculated in Fig.~\ref{fig_disprel_BLOB}, but also the time needed in the collision to generate a low-density phase where instabilities may develop, and the overall effect of the kinematics in an open system. 

\section{Dynamics of clusterisation in open systems, applications to heavy-ion collisions}\vspace{.75ex}

The composition of the two schemes presented at the beginning of Sec.~\ref{sec_NM}
and applied to open systems provides an analysis of the isovector and isoscalar properties of clustering in stable and unstable condition, in connection with analytic expectations from Fermi fluids.

\subsection{Spinodal clusters from density ripples at Fermi energies \label{sec_densityripples}}\vspace{.75ex}

%
%
\begin{figure}[h]
\includegraphics[width=1\textwidth]{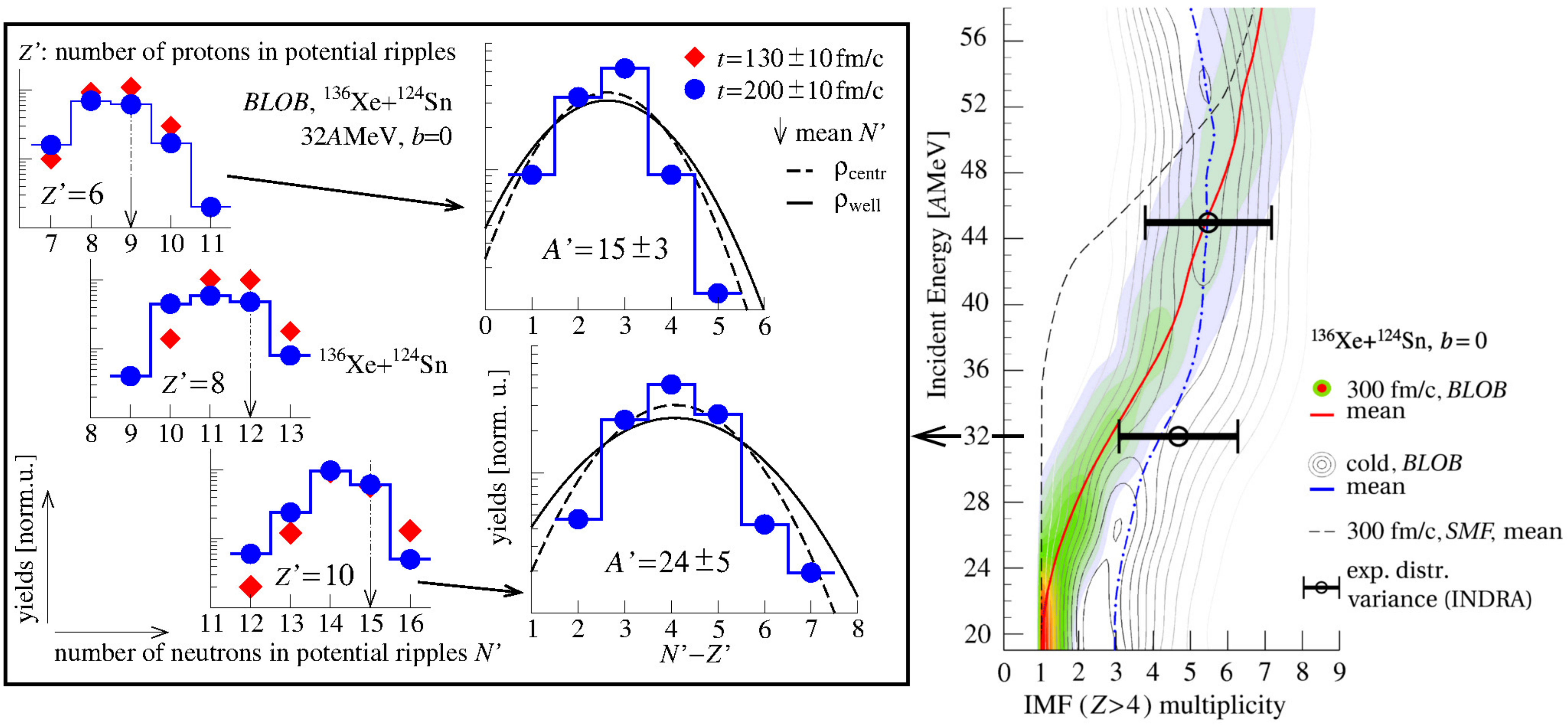}\\
\begin{minipage}[b]{1\textwidth}\caption{\label{fig_formingclusters}
	In the frame, BLOB calculation of the isotopic production in central collisions in $^{136}$Xe$+^{124}$Sn at 32~$A$MeV.
	(Left column in the frame) mass distributions of forming isotopes of C, O and Ne at early times $t=130$ and 200~fm/c, before separation into independent fragments.
	(Right column in the frame) Distributions of isotopic variances in potential ripples containing $N'$ neutrons and $Z'$ protons for the most probable configurations for forming clusters in the regions around C ($Z'=6$ and $A'=15$) and Ne ($Z'=10$ and $A'=24$) at $t=200$~fm/c.
	For comparison, analytic distributions are plotted, as a function of the density measured in the minimum   of the potential ripples $\rho_{\textrm{centr}}$ or averaged over the potential ripples $\rho_{\textrm{well}}$.
	The right panel shows the evolution of the multiplicity of fragments produced in central collisions in $^{136}$Xe$+^{124}$Sn at various incident energies, calculated with BLOB (colour map) at $t=300$~fm/c when fragments separate, and at the end of the whole decay sequence, calculated by BLOB+Simon (grey contours). The mean values and distribution widths are compared with experimental data from INDRA~\cite{Moisan2012,Ademard2014} (the bar indicates the variance of the multiplicity distribution).
}
\end{minipage}
\end{figure}
%

%
%
\begin{figure}[h]
\includegraphics[width=.525\textwidth]{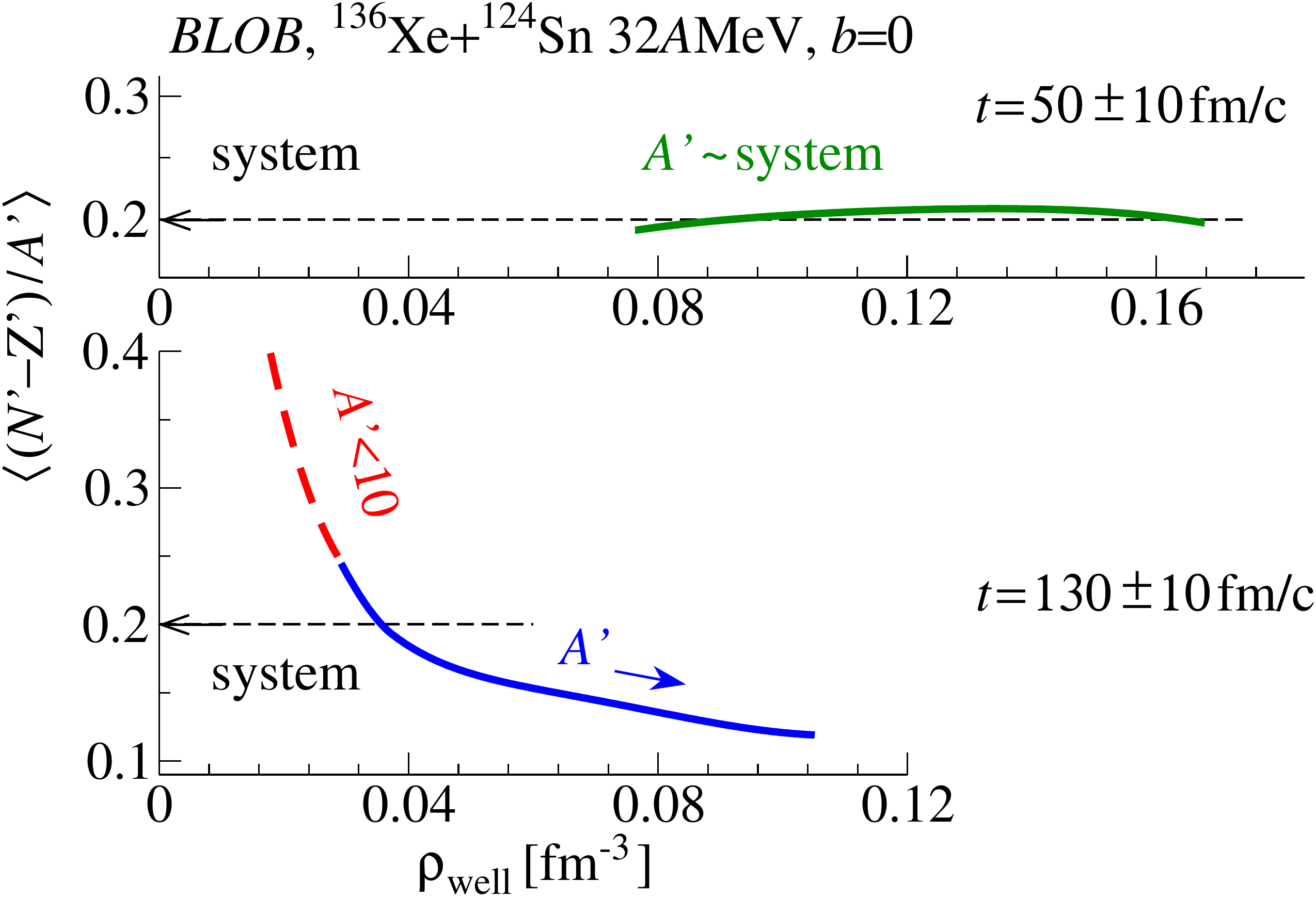}\hspace{2pc}%
\begin{minipage}[b]{.42\textwidth}\caption{\label{fig_distillation}
	The isotopic properties of fragments produced in central collisions in $^{136}$Xe$+^{124}$Sn at 32 $A$MeV is shown to evolve according to a process of isospin distillation.
	The average isospin content in potential ripples as a function of $\rho_{\textrm{well}}$ is calculated with BLOB before the onset of the fragmentation process at $t=50$fm/c, or when the process is almost achieved at $t=130$fm/c.
}
\end{minipage}
\end{figure}

An analysis which directly corresponds to the above calculations in nuclear matter is dedicated to the properties of emerging clusters in an open system at low density, produced in a heavy-ion collision.
Even before that clusters separate into well identified blobs of matter in configuration space, we can track the evolution of potential ripples developing in the bulk since early times.
In analogy to independent fragments, we can attribute to those inhomogeneities a mean radius and a density averaged over the corresponding potential well in order to obtain mass and element numbers $A',Z'$.
An indicative expectation for isotopic yield distributions may be obtained by analogy to eq.~(\ref{eq:fluctuation_dissipation}) as
\begin{equation}
	Y \approx \textrm{exp}[-(\delta^2/A')\,C_{\textrm{sym}}(\rho)/T]\;.
\label{eq:isotopic}
\end{equation}

These yields of forming fragments reflect the isovector fluctuations.
As already argued in Sec.~\ref{sec_iv}, while in equilibrated nuclear matter fluctuations are hard to revive and entertain due to low collision rates, in open systems fluctuations are initially built out of equilibrium, so that collision rates are larger.
	As a consequence, the isovector fluctuations variance results more consistent with analytic expectations, as illustrated in Fig.~\ref{fig_formingclusters} (in the frame) for the system $^{136}$Xe$+^{124}$Sn at 32 $A$MeV (central collisions), where the isotopic distribution of forming fragments is calculated with BLOB and compared to the expectation of eq.~(\ref{eq:isotopic}).

Fig.~\ref{fig_distillation} extends the study of the isospin content of forming clusters and fragments to the whole range of sizes produced in the process, and illustrates the corresponding average isospin content at a very early time ($t=50$fm/c), when the system is still rather homogeneous, and at a later time ($t=130$fm/c), when inhomogeneities have been built, as a function of the local density $\rho_{\textrm{well}}$ associated to the density ripple.
The larger neutron enrichment of the more volatile phase signs the onset of a distillation process~\cite{Chomaz2004}.

The right panel of Fig.~\ref{fig_formingclusters} extends the survey of fragment production to a large incident-energy range, examining the multiplicity of fragments with $Z>4$.
In this case, fragments are studied at $t=300$fm/c, when they are completely separated but still highly excited, and at the end of the full decay sequence (where the Simon transition-state model~\cite{Durand1992} was employed to complete the decay with sequential evaporation).
The BLOB+Simon calculation yields mean values and widths which are in agreement with experimental data from INDRA~\cite{Moisan2012,Ademard2014}.
The transition from fusion to multifragmentation occurs below 30$A$MeV and the dominant spinodal mechanism is gradually replaced by vaporisation beyond 45$A$MeV~\cite{Napolitani2013}.
	On the other hand, the simplified BL approach to introduce fluctuations used in SMF results less efficient in building fluctuations, and does not succeed in describing the transition from fusion to multifragmentation at the correct incident energy.

%

\subsection{Light systems}\vspace{.75ex}
%
%
\begin{figure}[b!]
\includegraphics[width=.48\textwidth]{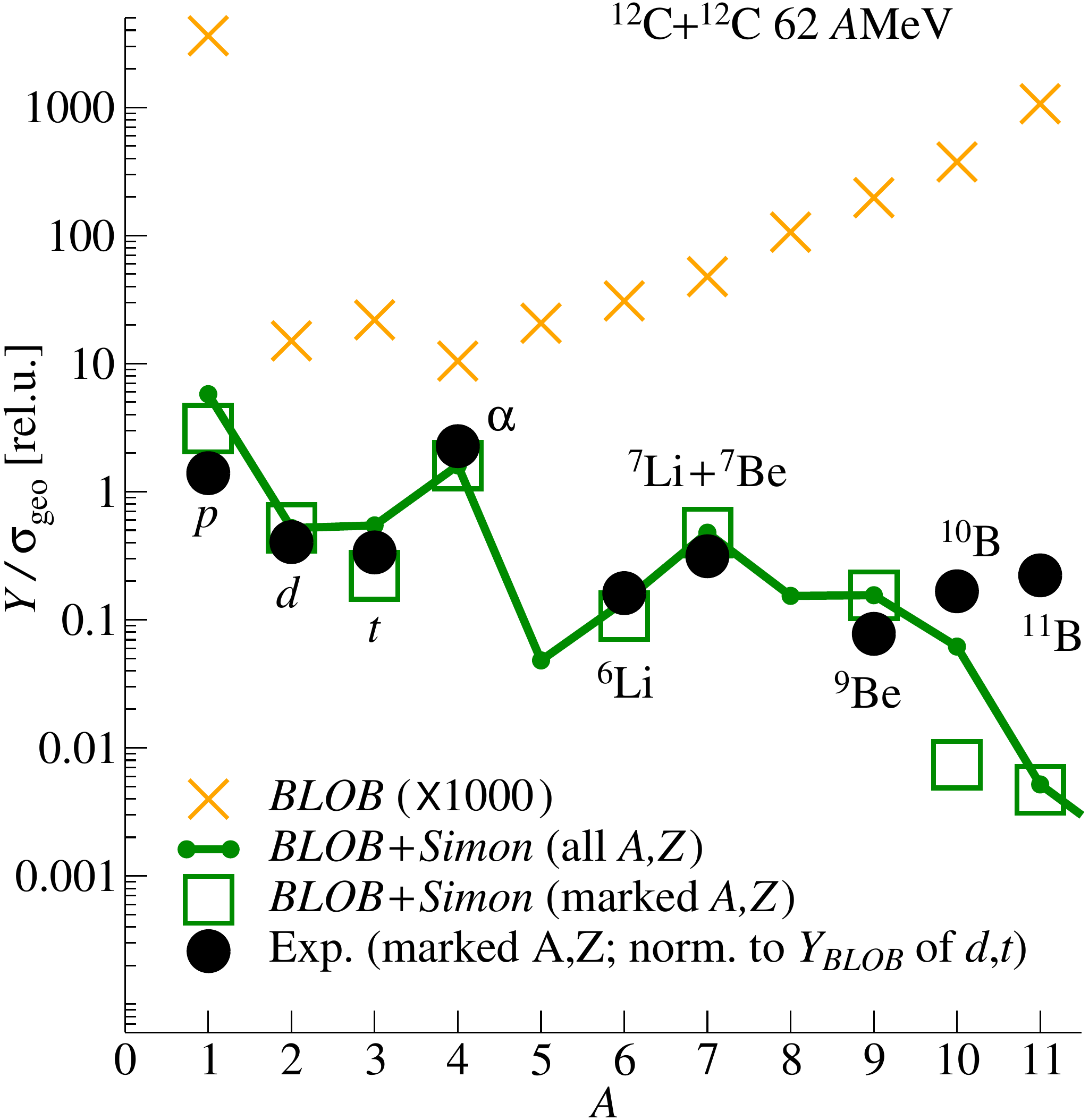}\hspace{2pc}
\begin{minipage}[b]{.425\textwidth}\caption{\label{fig_CC62b0to6}
	Clusters and fragment yields in $^{12}$C$+^{12}$C at 62~$A$MeV, divided by the geometric cross section, are shown as calculated with BLOB till the last registered split in a time window of 80 to 140~fm/c (BLOB, scaled by 1000) and after the full decay sequence, for which the model Simon~\cite{Durand1992} was employed (BLOB+Simon). A free $\sigma_\NN$ and an asy-stiff form for the symmetry energy are used.
	Experimental yields from ref.~\cite{DeNapoli2012} for some isotopes named on the plot, are indicated with dots; data are normalised so that the sum of $d$ and $t$ yields equals the corresponding calculated quantity. 
	Squares are calculations corresponding to the isotopes that have been measured.
}
\end{minipage}
\end{figure}

heavy-ion collisions and nuclear matter involve anditional clustering processes, where alpha and light charge particles are formed.
Such mechanism, different from dynamical instabilities, would require the explicit inclusion of further correlations in the present formalism.
	Light charged particles related to nuclear clustering have in fact too small size, exceeding the ultraviolet cutoff of the dispersion relation, so that they can not belong to the unstable multipole modes which characterise spinodal fragmentation. 
	Solutions for an explicit treatment of cluster formation are proposed in refs.~\cite{Danielewicz1991,Ono2016}.

	The ability of the stochastic approach of BLOB in handling fragment formation and introducing correlations bejond the level of kinetic equations can be considered a first approximation to light clusters, which are essentially emerging from dynamical fluctuations.
	Blob already takes charge of a fraction of the light cluster production and a sequential-decay afterburner (Simon~\cite{Durand1992} was used) adds the missing fraction from the decay of heavier elements.
	A comparison with data from ref.~\cite{DeNapoli2012} is presented in Fig.~\ref{fig_CC62b0to6} for the system $^{12}$C$+^{12}$C at 62~$A$MeV.

	At this level of comparison, the production rates of the light masses in the isobaric distribution seems consistently described.
	Such production is fed by the most central impact parameters and fades with periferal configurations, which are related to the heavier side of the mass distribution.


\subsection{Fluctuations and transparency at intermediate energy}\vspace{.75ex}
%
%
\begin{figure}[b!]
\includegraphics[width=1\textwidth]{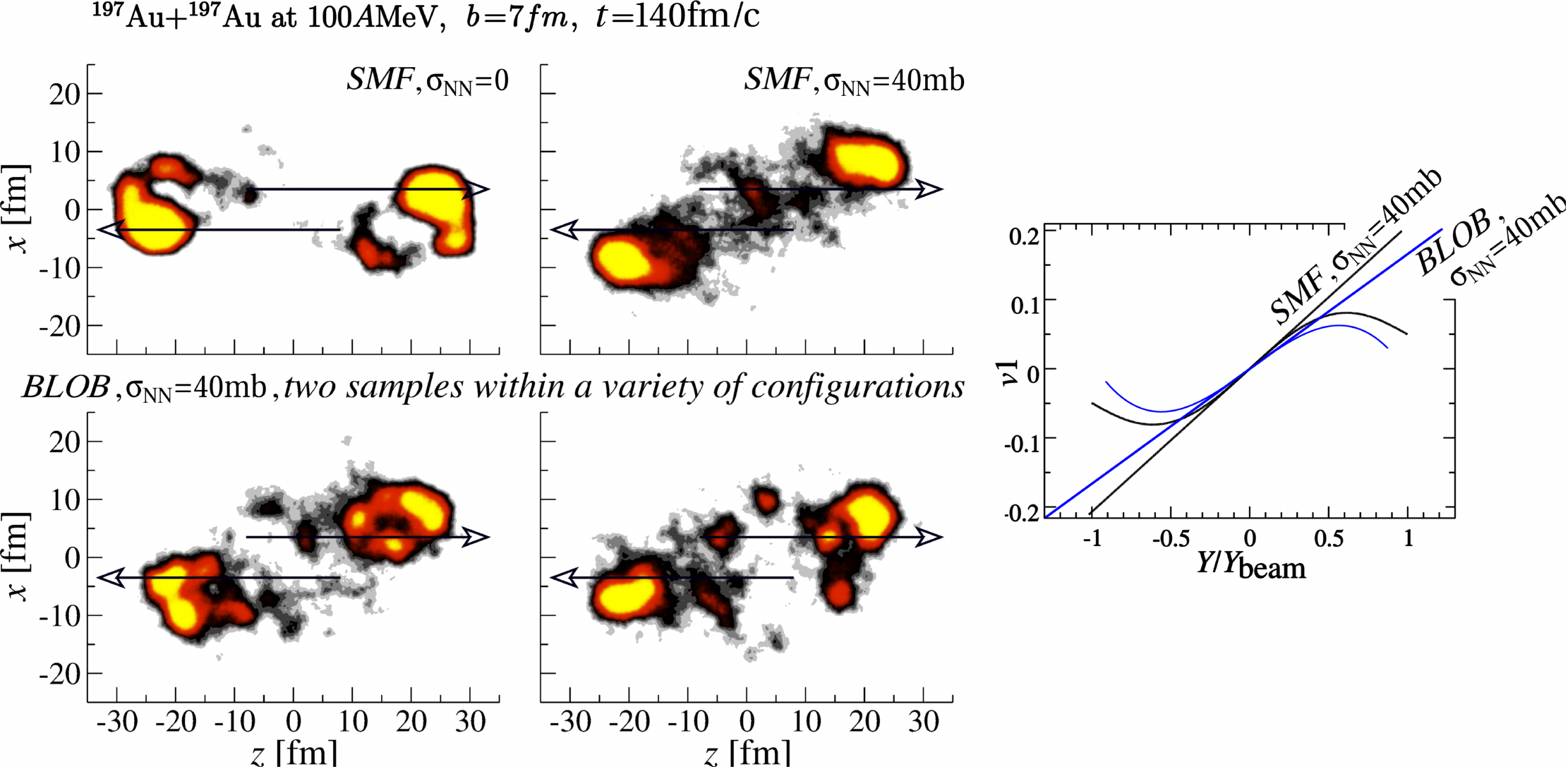}
\begin{minipage}[b]{1\textwidth}\caption{\label{fig_AuAu}
	Simulation of the collision $^{197}$Au$+^{197}$Au at $100\,A$MeV for an impact parameter of 7fm, studied at the time $t=140$~fm/c and simulated through three approaches: SMF without collision contribution (top, left in the density map), SMF (top, right in the density map) and BLOB (bottom row in the density map).
	The arrows indicate the direction of the target and projectile; their origins indicate the centres of target and projectile at the initial time $t=0$ for the simulation.
	(Right panel) Directed flow as a function of reduced rapidity for SMF and BLOB. The corresponding slope at zero reduced rapidity is indicated.
}
\end{minipage}
\end{figure}
	The connection between fragment production and hydrodynamic properties like the characterisation of the flow~\cite{ReisdorfRitter1997} at intermediate energies is strictly related to the treatment of fluctuations.

	At intermediate energy the inclusion of fluctuations has two antagonist effects: on the one hand, it enhances the fragmentation of the system, on the other hand it reduces the directed flow. 
	This effect can be studied in the comparisons of Fig.~\ref{fig_AuAu}, for the collision $^{197}$Au$+^{197}$Au at 100 $A$MeV for an impact parameter of 7fm.
	The simulation is performed with three approaches, SMF without and with a collision term (constant $\sigma_\NN=40$mb) and BLOB (also with$\sigma_\NN=40$mb) , using identical parameters for the mean field as defined in ref.~\cite{Jun2016}.
	The SMF approach describes the outward deflection of the trajectory imparted by the directed flow, which is absent in the SMF description without collision term.
	The BLOB approach exhibits a reduced directed flow with respect to SMF, because it competes with the production of fragments and clusters.
	This latter, due to the Langevin fluctuations, results in a large variety of very different fragment configurations; two of those are shown, one where the fragmentation of the quasi-target and the quasi-projectile is observed (bottom row in the density map, left), the other where the emitting source is situated at midrapidity (bottom row in the density map, right).

	A quantitative study of the flow is illustrated in the right panel of Fig.\ref{fig_AuAu}.
	For a simulation where the collision rate is identical (due to using the same constant $\sigma_\NN$), the larger fragmentation rate is reflected in a smaller slope for the directed flow as a function of reduced rapidity.
	This is a rather general example to spots the main difference between a simplified description of fluctuations and a BL approach solved in full phase space, when applied to intermediate energies.

\section{Conclusions}\vspace{.75ex}


The process leading a fermionic system, like nuclear matter or violent nuclear reactions, to separate into clusters and fragments is reviewed in some significant steps

Among more possible strategies to address the problem (see introduction), we have chosen to extend a mean-field description to include fluctuations in a Boltzmann-Langevin framework~\cite{Napolitani2017}
and suggested some scheme for testing the transport approach on analytic expectations.

The approach is particularly adapted to describe mean-field inhomogeneities, like the spinodal decomposition, but also conditions like Landau damping and other collective behaviours.
It reduces to an approximation when the formation of the light clusters should be described because no explicit cluster contributions are so far treated.
For larger clusters and intermediate-mass fragments in general, the approach yields quantitative predictions for the production of fragments in a broad range of situations and mechanisms encountered at Fermi energies~\cite{Napolitani2013, Colonna2017, Napolitani2016}.
To complete the survey, a further application to relativistic spallation has been undertaken elsewhere~\cite{Napolitani2015}.


\section*{References}\vspace{.75ex}


\begin{thebibliography}{9}
%
\bibitem{Pines1966}
Pines D and Nozieres P 
1966 \textit{The Theory of Quantum Liquids} (New York: Benjamin)
%
\bibitem{Napolitani2017}
Napolitani P and Colonna M 
\bbsty{\textit{Phys. Rev.} C}{96}{054609}{2017}
%
\bibitem{Ono2006_WCI}
Ono Akira and Randrup J 
\bbsty{\textit{Eur. Phys. J.}}{A30}{109}{2006}
%
\bibitem{Ono1992}
Ono Akira, Horiuchi H, Maruyama T and Ohnishi A
\bbsty{\textit{Phys. Rev. Lett.}}{68}{2898}{1992}; 
%
\bibitem{Ono2016}
Ono Akira
\bbsty{\textit{EPJ Web Conf.}}{122}{11001}{2016}
%
\bibitem{Feldmeier1990_1995}
Feldmeier H,
\bbsty{\textit{Nucl. Phys.}}{A515}{147}{1990};
%
Feldmeier H, Bieler K and Schnack J
\bbsty{\textit{Nucl. Phys.}}{A586}{493}{1995}
%
\bibitem{Balescu1976}
Balescu R
1976 \textit{Equilibrium and Non-equilibrium Statistical Mechanics} (New York: Wiley)
%
\bibitem{Ayik1988}
Ayik S and Gr\'egoire C
\bbsty{\textit{Phys. Lett. B}}{212}{269}{1988}
%
\bibitem{Lacombe2016}
Lacombe L, Suraud E, Reinhard P -G and Dinh P M
\bbsty{\textit{Ann. Phys.(NY)}}{373}{216}{2016}
%
\bibitem{Reinhard1992}
Reinhard P -G and Suraud E 
\bbsty{\textit{Ann. of Physics}}{216}{98}{1992}
%
\bibitem{Napolitani2013} 
Napolitani P and Colonna M 
\bbsty{\textit{Phys. Lett. B}}{726}{382}{2013}
%
\bibitem{Napolitani2012}
Napolitani P and Colonna M 
\bbsty{\textit{EPJ Web of Conf.}}{31}{00027}{2012}
%
\bibitem{Rizzo2008}
Rizzo J, Chomaz Ph and Colonna M
\bbsty{\textit{Nucl. Phys.}}{A806}{40}{2008}
%
\bibitem{Colonna1998}
Colonna M et al.
\bbsty{\textit{Nucl. Phys.}}{A642}{449}{1998}
%
\bibitem{Guarnera1996}
Guarnera A, Colonna M and Chomaz Ph.
\bbsty{\textit{Phys. Lett. B}}{373}{267}{1996}
%
\bibitem{Baran2005}
Baran V, Colonna M, Greco V and Di Toro M
\bbsty{\textit{Phys. Rep.}}{410}{335}{2005}
%
\bibitem{Colonna1993}
Colonna M, Burgio G F, Chomaz Ph, Di Toro M and Randrup J
\bbsty{\textit{Phys. Rev.} C}{47}{1395}{1993}
%
\bibitem{Colonna1994_a} 
Colonna M, Chomaz Ph and Randrup J
\bbsty{\textit{Nucl. Phys.}}{A567}{637}{1994}
%
\bibitem{Colonna2013}
Colonna M
\bbsty{\textit{Phys. Rev. Lett.}}{110 }{042701}{2013}
%
\bibitem{Larionov2000}
Larionov A B, Cabibbo M, Baran V and Di Toro M
\bbsty{\textit{Phys. Rev.} C}{61}{064614}{2000}
%
\bibitem{Kolomietz1996}
Kolomietz V M, Plujko V A and Shlomo S
\bbsty{\textit{Phys. Rev.} C}{54}{3014}{1996}
%
\bibitem{Landau1957} 
Landau L D
\bbsty{\textit{Sov. phys. JETP}}{5}{101}{1957}
%
\bibitem{Kalatnikov1958}
Kalatnikov I M and Abrikosov A A
\bbsty{\textit{Sov. phys. JETP}}{6}{84}{1958}
%
\bibitem{Chomaz2004}
Chomaz Ph, Colonna M and Randrup J
\bbsty{\textit{Phys. Rep.}}{389}{263}{2004}
%
\bibitem{Pomeranchuk1959}
Pomaranchuk I Ia
\bbsty{\textit{Sov. phys. JETP}}{8}{361}{1959}
%
\bibitem{Kolomirtz1999}
Kolomietz V M and Shlomo S
\bbsty{\textit{Phys. Rev.} C}{60}{044612}{1999}
%
\bibitem{Colonna1994}
Chomaz Ph and Colonna M
\bbsty{\textit{Phys. Rev} C}{49}{1908}{1994}
%
\bibitem{Napolitani2015} 
Napolitani P and Colonna M
\bbsty{\textit{Phys. Rev.} C}{92}{034607}{2015}; 
\bbsty{\textit{Il Nuovo Cimento}}{39 C}{384}{2016}
%
\bibitem{Durand1992} 
Durand D 
\bbsty{\textit{Nucl. Phys.}}{A541}{266}{1992}
%
\bibitem{Ademard2014}
Ademard G et al. (INDRA Coll.)
\bbsty{\textit{Eur. Phys. J. A}}{50}{33}{2014}
%
\bibitem{Moisan2012}
Gagnon-Moisan F et al. (INDRA coll.)
\bbsty{\textit{Phys. Rev C}}{86}{044617}{2012}
%
%
\bibitem{Danielewicz1991}
Danielewicz P and Bertsch G F
\bbsty{\textit{Nucl. Phys.}}{A533}{712}{1991}
%
\bibitem{DeNapoli2012}
De Napoli M 
\bbsty{\textit{Phys. Med. Biol.}}{57}{7651}{2012}
%
\bibitem{ReisdorfRitter1997}
Reisdorf W and Ritter H G
\bbsty{\textit{Annu. Rev. Nucl. Part. Sci.}}{47}{663}{1997}
%
\bibitem{Jun2016}
Xu Jun et al.
\bbsty{\textit{Phys. Rev.} C}{93}{044609}{2016}
%
\bibitem{Colonna2017}
Colonna M, Napolitani P and Baran V
2017 \textit{Mean-Field Instabilities and Cluster Formation in Nuclear Reactions}
in \textit{Nuclear Particle Correlations and Cluster Physics} 
(World Scientific), pp. 403–424
%
\bibitem{Napolitani2016}
Napolitani P and Colonna M
\bbsty{\textit{EPJ Web of Conf.}}{117}{07008}{2016}


\end{thebibliography}
\end{document}